\documentclass{article}

\usepackage{microtype}
\usepackage{graphicx}
\usepackage{subcaption}
\usepackage{booktabs} 
\usepackage{hyperref}

\usepackage[accepted]{icml2024}

\usepackage{amsmath}
\usepackage{amssymb}
\usepackage{mathtools}
\usepackage{amsthm}
\usepackage{bm}
\usepackage[capitalize,noabbrev]{cleveref}

\theoremstyle{plain}

\theoremstyle{definition}

\theoremstyle{remark}

\usepackage[textsize=tiny]{todonotes}

\icmltitlerunning{Inpainting Galaxy Counts onto N-Body Simulations over Multiple Cosmologies and Astrophysics}

\begin{document}

\twocolumn[
\icmltitle{Inpainting Galaxy Counts onto N-Body Simulations over Multiple Cosmologies and Astrophysics}

\icmlsetsymbol{equal}{*}

\begin{icmlauthorlist}
\icmlauthor{Antoine Bourdin}{McGill,Ciela,Mila}
\icmlauthor{Ronan Legin}{Ciela,Mila,UdeM}
\icmlauthor{Matthew Ho}{Sorbonne}
\icmlauthor{Alexandre Adam}{Ciela,Mila,UdeM}
\icmlauthor{Yashar Hezaveh}{Ciela,Mila,UdeM,CCA,Perimeter,TSI}
\icmlauthor{Laurence Perreault-Levasseur}{Ciela,Mila,UdeM,CCA,Perimeter,TSI}
\end{icmlauthorlist}

\icmlaffiliation{McGill}{Department of Physics, McGill University, Montréal, Canada}
\icmlaffiliation{Ciela}{Ciela Institute, Montréal, Canada}
\icmlaffiliation{Mila}{Mila - Quebec Artificial Intelligence Institute, Montréal, Canada}
\icmlaffiliation{UdeM}{Department of Physics, Université de Montréal, Montréal, Canada}
\icmlaffiliation{CCA}{Center for Computational Astrophysics, Flatiron Institute, New York, USA}
\icmlaffiliation{Perimeter}{Perimeter Institute for Theoretical Physics, Waterloo, Canada}
\icmlaffiliation{TSI}{Trottier Space Institute, McGill University, Montréal, Canada}
\icmlaffiliation{Sorbonne}{CNRS \& Sorbonne Universit\'e, Institut d'Astrophysique de Paris (IAP), Paris, France}

\icmlcorrespondingauthor{Antoine Bourdin}{antoine.bourdin@mail.mcgill.ca}
\icmlkeywords{Machine Learning, ICML}

\vskip 0.3in
]

\printAffiliationsAndNotice{} 

\begin{abstract}

Cosmological hydrodynamical simulations, while the current state-of-the art methodology for generating theoretical predictions for the large scale structures of the Universe, are among the most expensive simulation tools, requiring upwards of 100 millions CPU hours per simulation. N-body simulations, which exclusively model dark matter and its purely gravitational interactions, represent a less resource-intensive alternative, however, they do not model galaxies, and as such cannot directly be compared to observations. In this study, we use conditional score-based models to learn a mapping from N-body to hydrodynamical simulations, specifically from dark matter density fields to the observable distribution of galaxies. We demonstrate that our model is capable of generating galaxy fields statistically consistent with hydrodynamical simulations at a fraction of the computational cost, and demonstrate our emulator is significantly more precise than traditional emulators over the scales 0.36 $h\ \text{Mpc}^{-1}$ $\leq$ k $\leq$ 3.88 $h\ \text{Mpc}^{-1}$.

\end{abstract}

\section{Introduction} \label{sec:intro}

One of the most powerful tools in cosmology are hydrodynamical (hydro-) simulations, which are able to evolve the universe from billions of years into the past until present day \citep[e.g.][]{Illustris_TNG, Simba, Astrid}. These simulations make detailed and reliable predictions for the distribution of matter in the universe by carefully modeling the complex interactions of gravity, mass, thermodynamics, and electrodynamics. They have developed into a cornerstone of modern cosmology, creating a gateway connecting observations of the night sky with physical understandings of dark matter \citep{bozorgnia2017implications}, dark energy \citep{baldi2010hydrodynamical}, galaxies \citep{vogelsberger2020cosmological} and black holes \citep{chen2022dynamical}.

These powerful simulations come at a cost: they are extremely computationally expensive. High-resolution hydro-simulations can require hundreds of millions of CPU-hours to properly resolve galactic astrophysics \citep{vogelsberger2014introducing}. This computational cost is due to the $O(n^2)$ scaling of computing thermodynamic interactions between dark matter, gas, and star particles and is a major barrier for modern cosmology. Furthermore, recent works attempt to run not only one hydro-simulation, but suites of thousands \citep[e.g.][]{villaescusa2023camels, schaye2023flamingo}, with each simulation varying the configurations of cosmological models and galaxy astrophysics. These suites are essential to understand the effect of (latent) physical model parameters on astronomical observables, hence making it possible to infer the value of these latent parameters for our Universe from survey data, with traditional method but also using machine learning \citep[e.g.][]{de2023robust, kugel2023flamingo, ho2023benchmarks, ho2024ltu}. With the increasing quality and volumes of survey data, such suites of hydro-simulations are projected to drive the next-generation of computational cosmology \citep{national2021pathways}, and it is therefore paramount to develop faster methods for running them which exceed the traditional scaling laws.

A cheaper alternative to hydro-simulations are N-body simulations \citep[e.g.][]{heitmann2019outer, maksimova2021abacussummit, villaescusa2020quijote}, which only seek to model interactions between collision-less dark matter particles. Ignoring thermodynamic interactions reduces the computational scaling of N-body simulations to $O(n\log n)$, a marked improvement over hydro-simulations. N-body simulations are accurate approximations for the spatial distribution of all matter, as dark matter is believed to make up $\sim 85\%$ of the total matter budget in the universe. However, as N-body simulations do not directly simulate stars and galaxies, we must rely on a plethora of simplifying assumptions when comparing them to real observations \citep{desjacques2018large}. The parameters of these assumptions are particularly challenging to constrain for populations of smaller galaxies which dominate the statistics of observational surveys \citep[e.g.][]{richardson2012halo, yuan2022abacushod, garcia2024hod}. In these high-resolution regimes, the use of hydro-simulations, or their respective emulators, is crucial.

In this work, we develop an emulator for high-resolution hydro-simulations by populating N-body simulations with galaxies. We make use of a conditional score-based diffusion model \cite{song2021scorebased, Ronan_IC} that generates samples containing the number of galaxies at different grid positions given the mass of matter evolved in an N-body simulation over a pixelated grid. 

Our model is trained on the 3D CAMELS-Astrid simulation from the LH set \cite{CMD}, which consists of one thousand pairs of N-body and hydro-simulations. Each pair of simulations has a different configuration of cosmological and astrophysical parameters, varying the total matter density ($\Omega_m$), the amplitude of matter density fluctuations ($\sigma_8$), the energy of galactic winds ($A_{\rm SN1}$), the wind speed of galactic winds ($A_{\rm SN2}$), the rate of energy injection from active galactic nuclei ($A_{\rm AGN1}$), and the radio mode threshold for active galactic nuclei ($A_{\rm AGN2}$). Each of these parameters has a strong, non-linear effect on the spatial distribution of galaxies in the universe.

Using a diffusion model with this dataset allows us to learn the distribution of galaxy counts over multiple cosmologies. This means that, given an N-body simulation at a fixed cosmology, we can generate in parallel 100 different realizations of galaxy counts which match the summary statistics of a hydro-simulation in 20 minutes. Meanwhile, obtaining an analogous result with traditional methods would entail running one hundred hydro-simulations, each requiring millions of CPU hours. 

Furthermore, we show that our model is able to correctly predict galaxy counts on cosmological and astrophysical parameters that were not seen during training, enabling us to explore how the parameter space of configurations affects galaxy clustering in a time-efficient manner. To the best of our knowledge, this is the first work connecting probabilistically N-body simulations to galaxies with the degree of accuracy of a hydro-simulations over a wide range of cosmological and astrophysical parameter values.


\section{Related Works} \label{sec:related_works}
Within this section we briefly discuss alternative works aiming to connect N-body simulation to baryonic properties. In section \ref{sub_sec:HOD}, we present the most commonly used closed-form parametric model used in cosmology while in section \ref{sub_sec:other_ML} we highlight other recent approaches which use machine learning. We highlight some of the advantages and limitations of each model. 

\subsection{Halo Occupancy Distribution} \label{sub_sec:HOD}

To circumvent the need for expensive hydrodynamical simulations, Halo Occupancy Distribution (HOD) models are often used to connect N-body simulations to galaxies \cite{Zheng_2005, Zheng_2007}. As the name suggests, these models identify halos which are clustering of dark matter over a certain length scale. Halos create gravitational potential wells which attract baryons and becomes locations where galaxies form. 

Once dark matter halos are identified in an N-body simulation, individual halos are assigned a probability of hosting a galaxy given their mass. The HOD model separates galaxies into two categories: a central galaxy sitting at the minima of the gravitational well and satellite galaxies which are progressively added and that orbit around the central galaxy. The expected value for the number of central ($N_{ \text{cen} }$) and satellite ($N_{ \text{sat}}$) galaxy as a function of halo mass ($M$) is described by, 
\begin{equation}\label{eq:basic_HOD}
    \begin{gathered}
        \left< N_{ \text{cen} }(M) \right>=
        \frac{1}{2} \left[ 
        1+\text{erf} \left( \frac{ \log{M}-\log{M_{ \text{min} }} }
        { \sigma_{ \text{logM} } } \right)
        \right]       
        \\
        \\
        \left< N_{ \text{sat} }(M) \right>=
        \left< N_{ \text{cen} }(M) \right> \left( \frac{ M - M_0 }{ M_1 } \right)^{\alpha}.
    \end{gathered}
\end{equation}
Here, $\text{erf}$ is the Gaussian error function. The parameter $M_{\text{min}}$ is the halo mass at which there is a 0.5 expected value of hosting a central galaxy. On the other hand, $\sigma_{ \text{logM} }$ controls how fast this expectation grows to one for different halo masses.  Similarly, $M_0$ determines the minimum halo mass to have a non-zero expectation of hosting a satellite galaxy. The parameter $M_1$ and $\alpha$ control how fast this expectation grows. We note that here the expectation is not bounded by one as we can have multiple satellite galaxies within a single halo. 

An HOD works by identifying all of the halos in a N-body along with their position and mass. It parses through each individual halo and computes the expected number of galaxies based on equation (\ref{eq:basic_HOD}). For central satellites, a Bernoulli distribution is sampled with the expectation value set by $\left< N_{ \text{cen} }(M) \right>$ to determine whether that halo hosts a central satellite. If so, a Poisson distribution with the expectation  
$\left< N_{ \text{sat} }(M) \right>$ is sampled to populate the halo with satellites. 

Because of their simplicity, HOD models are very inexpensive, and also have the notable advantage that they produce population of galaxies in a probabilistic manner. However, previous works have shown that simple HODs broadly marginalize over complex baryonic physics, thus lacking sufficient precision to match the data quality that is expected from upcoming observational surveys \cite{lim_HOD}. Furthermore, the HOD model described above ignores the effect of environment, which is known to significantly impact the process of galaxy formation. Finally, the parametrization in equation (\ref{eq:basic_HOD}) does not include cosmological parameters.  

\subsection{Alternative ML Approaches} \label{sub_sec:other_ML}

In prior work, machine learning techniques have been employed to estimate galaxy counts from N-body only simulations. For instance, \cite{Shirley_DM_gal} utilized convolutional neural networks to map from N-body simulations to galaxy counts, achieving superior accuracy compared to the traditional Halo Occupation Distribution (HOD). It is presumed these improved results were due to the fact that CNNs capture local information, hence, focusing on environmental factors instead of only considering halo mass as is done in an HOD. Nevertheless, a significant limitation of this approach is the deterministic nature of the model's predictions concerning galaxy numbers. Such deterministic predictions fail to encapsulate the inherent stochasticity of the task at hand, since the information from dark matter N-body simulations is insufficient to exactly predict galaxy counts from hydrodynamical simulations. 

Score-based generative models have shown significant promise in recent field-level inference tasks, where samples can be generated from learned probability distributions over cosmological fields \cite{Ronan_IC}. These models can address the limitations inherent in deterministic approaches as noted by \cite{Shirley_DM_gal}, while simultaneously providing a more realistic representation of galaxy counts compared to the Halo Occupation Distribution (HOD) model. Similarly, \citet{ono2024debiasing} used score-based generative models to infer the dark matter distribution from the galaxy field, using two-dimensional slices of hydro-simulations. While this approach was applied to a task opposite to that of the generative model under consideration here, this work demonstrated  that score-based generation could effectively capture the dark-matter-galaxy connection.

Motivated by these previous points, we explore the use of score-based generative modeling to develop a more accurate model for predicting galaxy counts from dark matter N-body simulations. Furthermore, we demonstrate that the score-based model can also learn to implicitly marginalize over cosmological parameters and improve accuracy over multiple cosmologies. 
A similar methodology was taken by \citet{cuesta2023point} which used diffusive graph neural networks to model the spatial distribution of halos in dark-matter-only simulations. While aptly capturing the uncertainty of generative modeling, this approach was limited to modeling the dark-matter-only component of simulations and also required complex graph representations, which greatly limited the generative capacity to only 5,000 halos under GPU memory limitations. Our chosen approach using convolutional architectures allows for simple factorization of the generative distribution over the box volume, allowing for effective `outpainting' to arbitrarily large volumes \citep[e.g.][]{rouhiainen2023super}.

In the following section, we provide a concise overview of score-based generative models and their application in sampling the probability distribution $p(\bm{x}|\bm{y})$ of plausible realizations of galaxy count fields $\bm{x}$ conditioned on dark matter N-body simulations $\bm{y}$.

\section{Score-Based Models}

Score-based generative modeling leverages neural networks to approximate the score - the gradient of the log probability of the data - which enables efficient sampling in high dimensional space from learned distributions. Specifically, given a data distribution $p(\bm{x})$, a neural network is trained to approximate the score of the distribution $\nabla_{\bm{x}} \log p(\bm{x})$. Typically, the goal of score-based generative modeling is to generate samples from a data distribution $p(\bm{x})$ by solving the following reverse-diffusion stochastic differential equation \cite{ANDERSON1982313, song2021scorebased},
\begin{equation}\label{rvdsde_cond}
    d\bm{x} = \left(f(\bm{x}, t) - g(t)^2 \nabla_{\bm{x}} \log p_{t}(\bm{x}) \right)dt + g(t) d\bm{w},
\end{equation}
where $\nabla_{\bm{x}} \log p_{t}(\bm{x})$ is the probability distribution of $\bm{x}$ at time $t$. To solve equation (\ref{rvdsde_cond}), a neural network $s(\bm{x}, t)$ is trained via denoising score matching \citep{Vincent_DSM, song2021scorebased, Hivarinen2005} to approximate the correct score of the data distribution $\nabla_{\bm{x}} \log p_{t}(\bm{x})$ for different levels of noise parameterized by a time variable $t$.

The data generation process achieved by solving equation (\ref{rvdsde_cond}) can be extended to sampling conditional distributions $p(\bm{x}|\bm{y})$ by training a neural network to approximate the conditional score $\nabla_{\bm{x}} \log p_{t}(\bm{x}|\bm{y})$. In this case, the neural network is fed the conditional variables $\bm{y}$ as additional inputs.

In this work, we train a neural network, denoted as $s(\bm{x}, \bm{y}, t)$, to learn the score of the conditional distribution $p_{t}(\bm{x}|\bm{y})$ of galaxy count fields $\bm{x}$ obtained from computationally expensive hydro-simulations conditioned on computationally inexpensive N-body simulations $\bm{y}$. By learning this distribution, we can solve equation (\ref{rvdsde_cond}) to efficiently generate realizations of galaxy count fields, and circumvent the computational cost of running direct hydro-simulations.

For sampling, we choose to use the Variance-Exploding Stochastic Differential Equation, which is characterized by $f(\bm{x}, t) = 0$ and $g(t) = \sqrt{\frac{d [\sigma^2(t)]}{dt}}$.  The noise level during sampling is set by $\sigma(t) = \sigma_{\text{min}} \left(\frac{\sigma_{\text{max}}}{\sigma_{\text{min}}} \right)^t$. We explain how the parameters ${\sigma_{\text{min}}}$ and ${\sigma_{\text{max}}}$ are set in section \ref{sec:discrete_diff}.

\subsection{Network Architecture}

Following the work of \citep{Ronan_IC} we adopt a U-net architecture \citep{Ronnenberger2015} with 3D convolutions to predict the conditional score distribution $p(\bm{x}|\bm{y})$. We closely follow the architecture code implemented in the package \texttt{score-models}\footnote{\href{https://github.com/AlexandreAdam/score_models}{github.com/AlexandreAdam/score\_models}}, specifically, our model contains 4 downsampling/upsampling levels, each featuring 2 residual blocks derived from the \texttt{BigGAN} model \citep{biggan}. The base number of feature maps is 32, and is doubled on the third and fourth downsampling step. We train the network for 50,000 epochs, with a batch size of 20, on a single NVIDIA A100 40GB GPUs. The network weights are optimized with the Adam optimizer \citep{Kingma2015} with a fixed learning rate of $2\times 10^{-4}$. We apply gradient clipping to restrict the weight gradients to a maximum norm of 1. The entire duration of training is approximately 70 hours for 2.25 million iterations.

To incorporate the condition variable $\bm{y}$, we introduce it as an additional channel in the network's input. Since the values in N-body simulations are to the power of 10 we take a base-10 logarithmic in order to have values ranging from roughly 12 to 16 allowing for training to be more stable. Since we do not directly input the cosmological and astrophysical parameters to the neural network, it learns the score of the conditional galaxy count distribution marginalized over these parameters.

\subsection{Diffusion and Discrete Data}\label{sec:discrete_diff}

Our goal is to learn the number of galaxies at each voxel which is an inherently discrete value. In recent years, frameworks for discrete diffusion have been proposed  \citep{santos2023blackout, austin2023structured, campbell2022continuous}; however their scalability and ability to perform as well as their continuous counterpart is still unclear. Therefore, we preferred developing our work with the well established continuous space denoising model. 

However, we observed that directly applying continuous space models to discrete data led to inaccurate results specifically being unable to predict high valued galaxy counts in the hundreds. To mediate this issue we decided to log our galaxy counts on a base-10 which made the galaxy counts range within zero and three. 

Based on this transformed data, we set the noising parameters $\sigma_{\text{max}}$ to the largest euclidean distance in the training set as proposed by \citep{Improve_yang_song}, which gave $\sigma_{\text{max}} = 44$. As for $\sigma_{\text{min}}$ it is set to $1\times 10^{-5}$ which is below the integer precision needed to identify individual galaxy counts when reversing the log transform. Based on these two parameters, we set the number of discretized steps needed to sample between t$\in$[1,0] when solving equation (\ref{rvdsde_cond}) to 3000 in order to have a 96\% change of staying in distribution  \citep{Improve_yang_song}. 

Furthermore, we follow the idea of passing high frequency information to our network in order to better learn small scale information and better predict values of individual voxels \citep{Kingma2021}. To do so we pass our logged galaxy data through a sin and cos function which have a set frequency of 64 and 128. These frequencies were seen to give the best results in \citep{Kingma2021}, and these four additional maps are given as extra channels to the network. This procedure helped in better modeling sudden low to high value changes between nearby voxels. 

\section{Simulations}
We used the Latin Hypercube (LH) set of simulations from the CAMELS-ASTRID suite \citep{CMD}. This dataset includes 1000 pairs of N-body and hydrodynamical simulations, each at a volume of 25 $(\text{Mpc}/h)^3$ and run using unique configuration of cosmological and astrophysical parameters. These parameters span a uniform prior over specified ranges: $\Omega_M \in [0.1, 0.5]$, $\sigma_{\text{8}} \in [0.6, 1.0]$, $A_{\rm SN1} \in [0.25,4]$, $A_{\rm SN2} \in [0.5,2]$, $A_{\rm AGN1}\in [0.25, 4]$, $A_{\rm AGN2}\in [0.25, 4]$.

We identify galaxies in the hydro-simulation by using the provided catalogue of halos and subhalos which were found with the Friends of Friends \citep{davis1985evolution} and SubFind \citep{springel2001populating} methods respectively. The chosen criteria for a subhalo to host a galaxy is containing stellar mass and having a mass greater than $10^{9.5}$ \(M_\odot\)$/h$ . We then place the galaxy catalogue on a 3D grid of size $32^3$. Similarly, for the N-body simulation, the dark matter mass is set on a 3D grid of size $32^3$. We divide the pair of gridded simulation into training and testing sets, allocating 900 simulations for training and 100 simulations for testing.

\section{Results}

\begin{figure*}[h!]
   \centering
    \begin{tikzpicture}
        \node at (0,0) {\includegraphics[scale = 0.5]{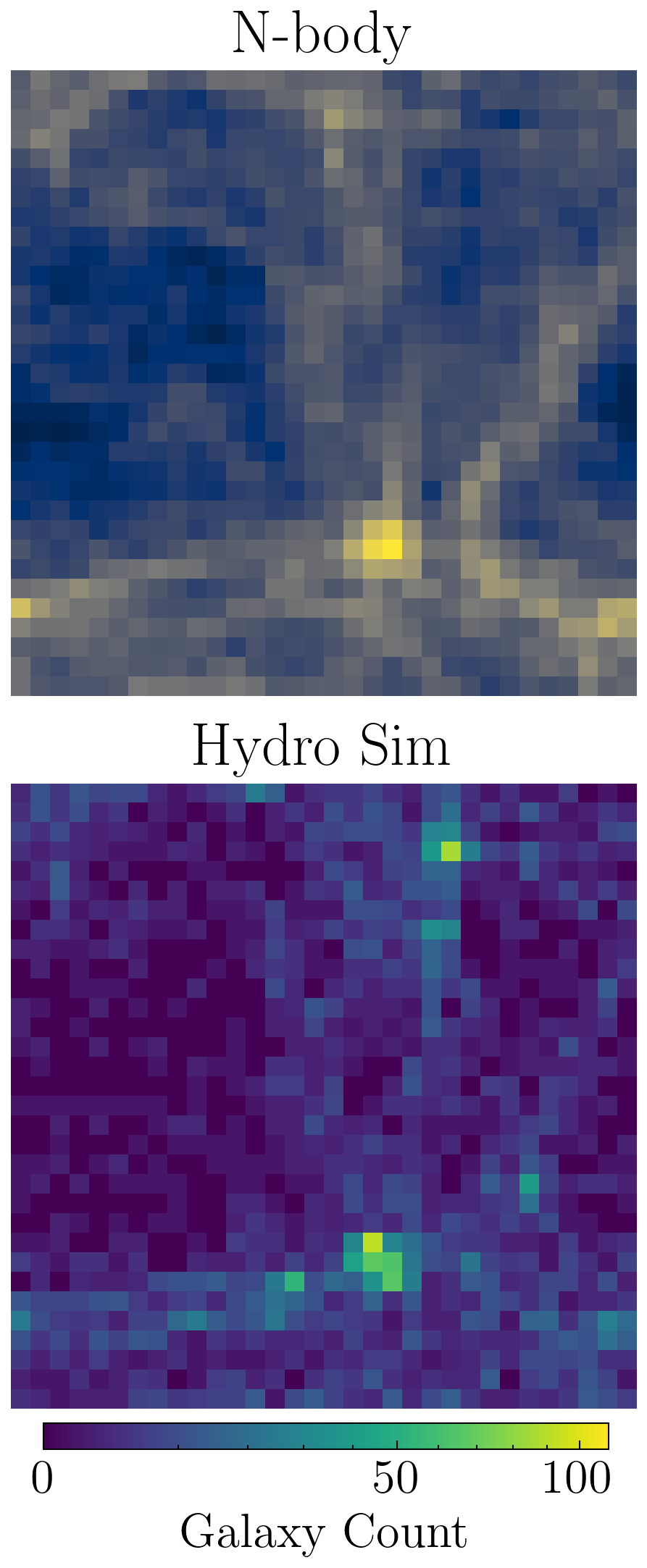}};
        \node at (8.88,0) {\includegraphics[scale = 0.5]{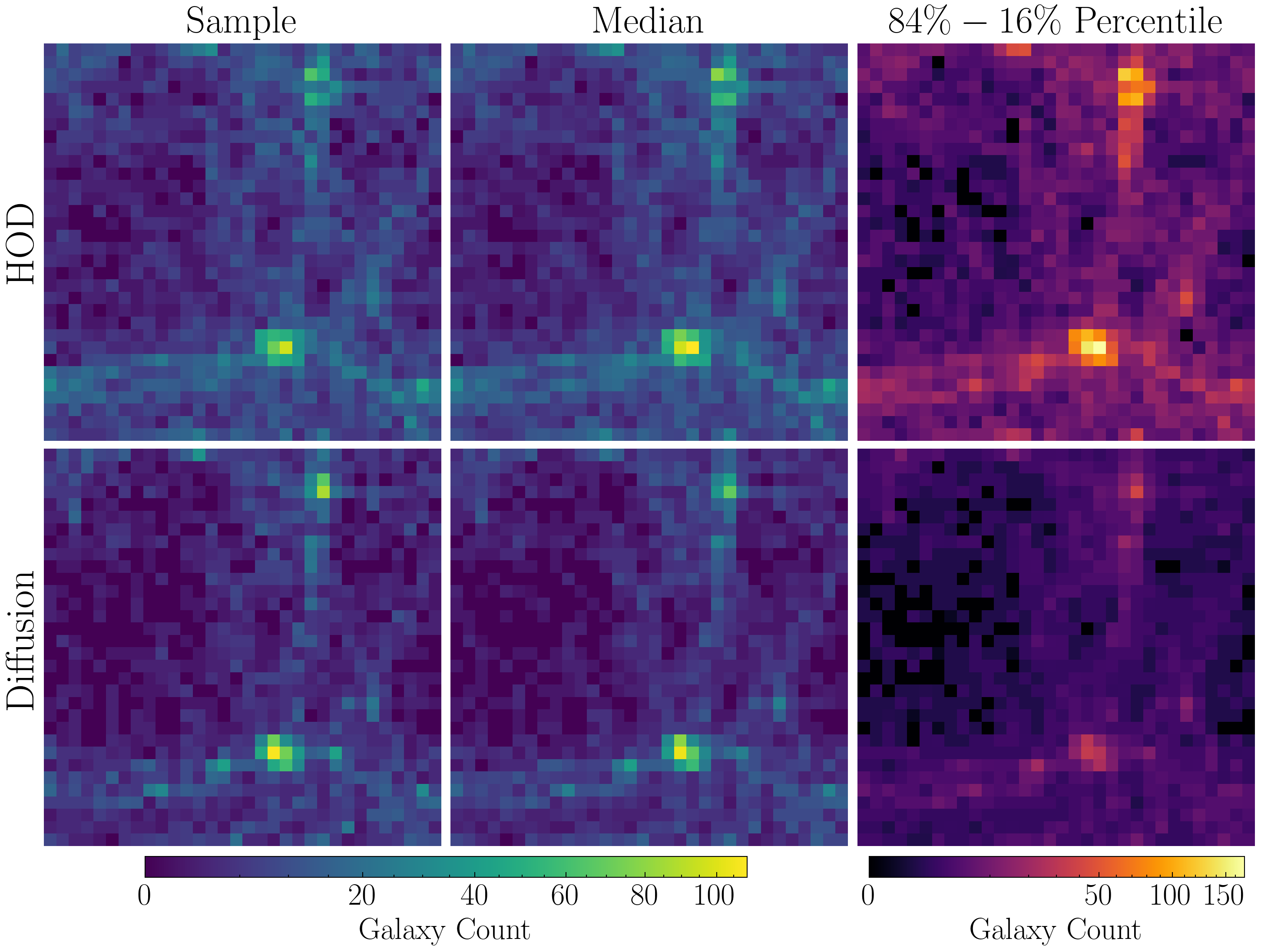}}; 
        \draw[black, thick] (2.38,-4.4) -- (2.38,4.45);
    \end{tikzpicture}
    \caption{Comparison of galaxy count fields from a hydrodynamical simulation with predictions made by the Halo Occupation Distribution (HOD) method and our score-based generative diffusion model based on the corresponding dark matter N-body simulation. Each of the 2D fields showcased were obtained by summing the fields over the depth of the full 3D volume. The parameters used for the hydro-simulation are $\Omega_M = 0.25140$, $\sigma_{\text{8}} = 0.8606$, $A_{\rm SN1} = 0.73002$, $A_{\rm AGN1} = 1.13131$, $A_{\rm SN2} =0.70466$, $A_{\rm AGN2} = 1.07848$. The leftmost column shows the dark-matter-only N-body and corresponding hydro-simulation, ran at the same cosmology and initial conditions. We then show the median and 84th-16th percentile range of the galaxy counts for both the HOD and diffusion models. The N-body is shown for illustrative purposes and its color bar is not presented. The 84th-16th percentile range has a different color bar to better reflect the spread of galaxy counts predictions made by the HOD. This comparison highlights the differences in spatial distributions of galaxy counts between the two models and the improved prediction of the diffusion model over HOD.}
    \label{visual_plot}
\end{figure*}

\begin{figure}[h!]
    \centering
    \includegraphics [width=\linewidth]{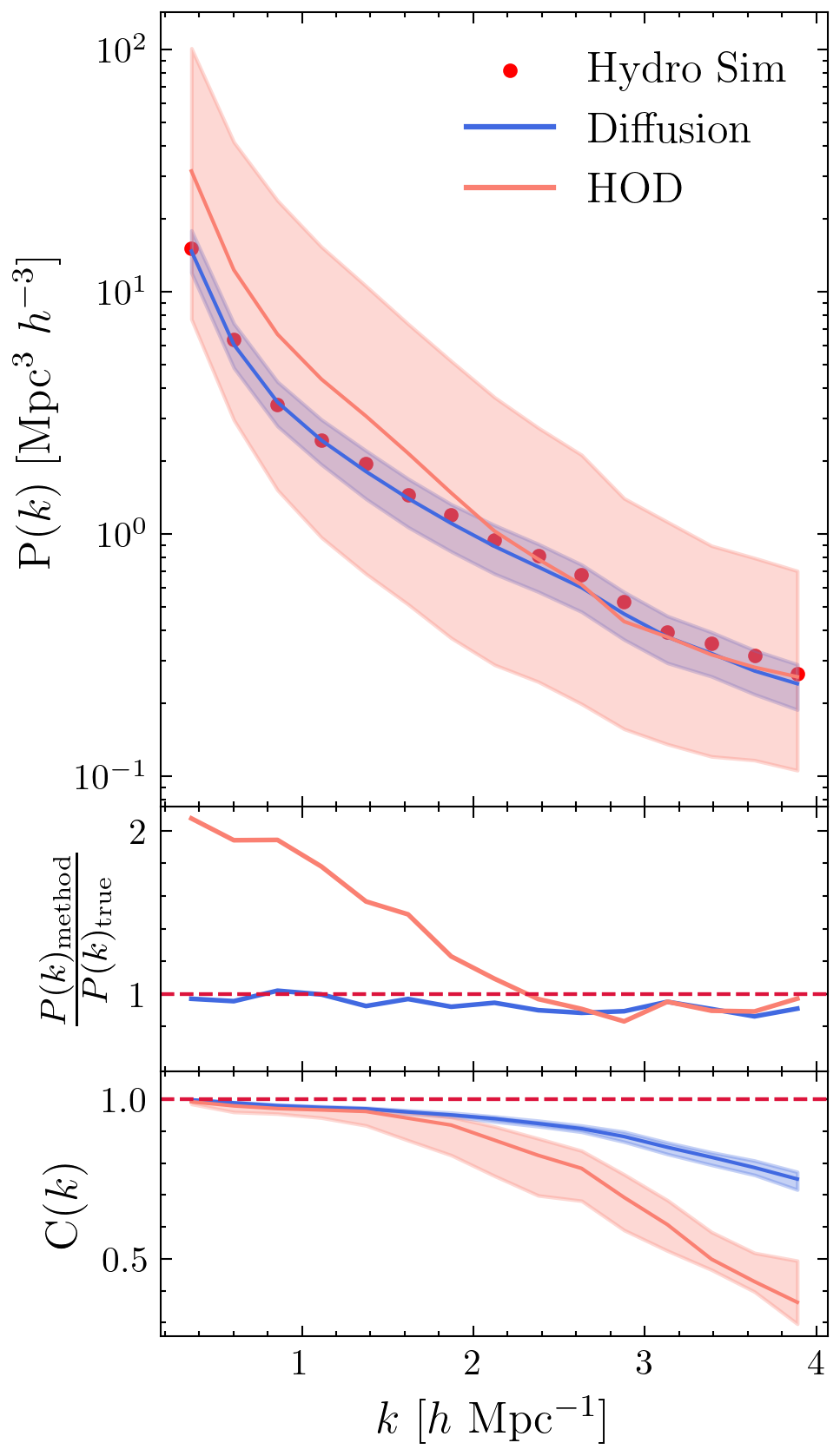}
    \caption{Summary statistics of the galaxy count fields for the results presented in Figure \ref{visual_plot}. The ground truth is the hydro-simulation $\bm{x}_{\text{true}}$ in red while conditional samples $p(\bm{x}|\bm{y})$ generated using the score network and the HOD are presented in blue and orange respectively. From top to bottom, for each method we plot the power spectra, power spectra ratio between the ground truth and generated samples, and the cross-correlation between the ground truth and generated samples. The median of the power and cross-correlation is presented for the diffusion and HOD while the uncertainty region is traced by the 16th and 84th percentile using 100 samples for both methods. }
    \label{power_plot}
\end{figure}

\begin{figure}[h!]
    \centering
    \includegraphics [width=\linewidth]{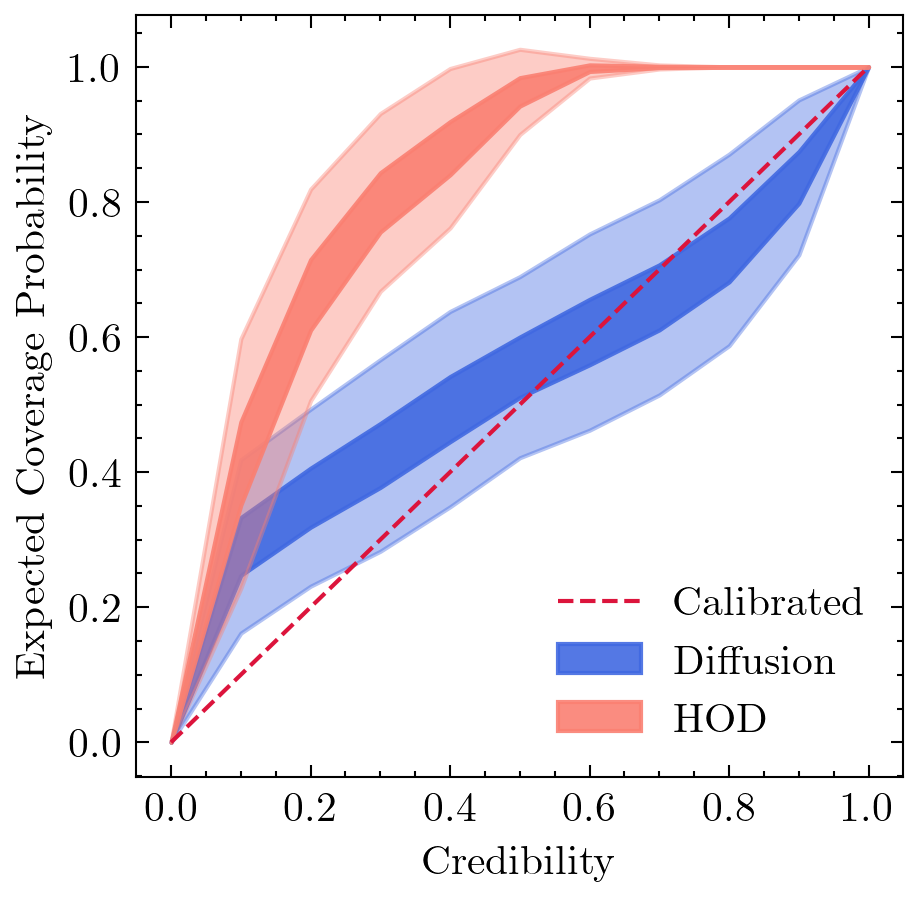}
    \caption{Coverage probability test using the TARP method to assess the accuracy of power spectra from samples generated by the diffusion model and the HOD method over the 100 hydro-simulations from the test set. The blue and orange curves represent the results for the diffusion model and HOD, respectively, with shaded areas indicating the $1\sigma$ and $3\sigma$ error obtained by bootstrapping examples from the test set. The dashed red line indicates the ideal calibrated line signifying accurate power spectra predictions. The diffusion model results show improved alignment with the diagonal compared to the HOD method, which exhibits a substantial deviation, indicating less accurate coverage.}
    \label{tarp_plot}
\end{figure}

For a given dark matter N-body simulation $\bm{y}$, we generate 100 samples of galaxy counts $\bm{x}$ from the conditional distribution $p(\bm{x}|\bm{y})$. This is achieved by solving equation (\ref{rvdsde_cond}) using our trained score network $s(\bm{x}, \bm{y}, t)$. Our testing set contains 100 different N-body simulations leading to a total of 10,000 samples generated using 20 NVIDIA A100 40GB GPUs in parallel within 80 minutes.  

In order to assess the validity of our samples we compare them to the hydro-simulation associated with the N-body simulation and the traditional HOD method. We take a Bayesian approach to best fit the parameters of the HOD to the power spectra of the fields over multiple cosmologies. Uniform priors are set on all 5 parameters of the HOD within the ranges $\log M_{\text{min}} \in [9, 12] $, $\sigma_{ \text{logM} } \in [0.1, 0.6]$, $\log M_0 \in [11, 16]$, $\log M_1 \in [11, 16]$ and $\alpha \in [0, 1.5]$. An analytic likelihood for the power spectra at each cosmology cannot be used for the CAMELS-Astrid LH dataset as it only has one simulation per cosmology. Therefore, a Neural Likelihood Estimator (NLE) \citep{SNLE} is trained on the power spectras of the 900 hydro-simulations in the training set. This allows the NLE to learn the conditional probability of HOD parameters which best reproduces the hydro-simulation power spectra over the different cosmologies in the training set. Finally, with the uniform priors and trained NLE, one hundred parameters from the posterior distribution are sampled using Markov Chain Monte Carlo at each cosmology in the testing set. 

We first present the results for a single N-body simulation at a fixed cosmology in the test set. Figure \ref{visual_plot} displays samples derived from the learned distribution alongside those generated by the HOD parametric method, accompanied by the corresponding true simulation of the galaxy count field. In Figure \ref{power_plot}, we present a comparison of summary statistics of the generated galaxy count samples versus the ground truth galaxy count field from the hydrodynamical simulation showcased in Figure \ref{visual_plot} (see additional examples of power spectra obtained for varying cosmological parameters in Appendix~\ref{sec:appendixA}). This comparison specifically includes the power spectrum, the transfer function (defined as the ratio of power spectra between the generated samples and the ground truth galaxy count field), and the cross-correlation of the two fields. 

In order to assess the accuracy of the learned conditional distribution $p(\bm{x}|\bm{y})$ over all cosmologies in the testing set, we employ the Test of Accuracy with Random Points (TARP) coverage test from \citet{lemos2023sampling}. This test determines whether the set of generated samples $\bm{x}_{\text{samples}}$ accurately reflects the true underlying probability distribution $p(\bm{x}|\bm{y})$ by measuring the coverage probability of the samples over randomly-generated credible regions. We perform a TARP test on the power spectra of the fields with the results being depicted in Figure \ref{tarp_plot}. The optimal case is a straight diagonal line indicating samples being generated within the underlying distribution formed by the hydro-simulation.

\section{Discussion}

Our score-based generative model addresses the limitations of previous methods for predicting galaxy count fields from N-body simulations by learning the score $\nabla \log p_t(\bm{x}|\bm{y})$ of the conditional probability distribution of galaxy counts $\bm{x}$ and employing a reverse-diffusion process to generate samples from this distribution conditioned on dark matter N-body simulation $\bm{y}$. The model is benchmarked against the halo occupation distribution (HOD) model, which also employs probabilistic sampling but relies on significant assumptions about the distribution of galaxy counts.

A critical shortcoming of the HOD method is its failure to incorporate information about the local environment. In contrast, the U-net architecture used for learning the score of the distribution mitigates this by analyzing the field at multiple resolution levels, and is therefore capable of extracting information from a wider spatial region. Additionally, HOD-based predictions are not dependent on cosmological or astrophysical parameter values, whereas in principle, score-based generative models are capable of extracting and marginalizing over this information from dark matter N-body simulations. This incorporation of local environmental factors and cosmological parameters renders score-based generative models as a more flexible approach for generating accurate galaxy count fields. A demonstration of this can be seen from Figure \ref{visual_plot}, where the fields generated by HOD appear to struggle to predict galaxy count features in regions of low density. Furthermore, it can be seen from Figure \ref{visual_plot} that HOD predicts an overly high galaxy counts in certain regions of high density. In comparison, the samples generated with our diffusion model are capable of reliably replicating features which are consistent with the hydro-simulation in both low and high density regions. Additionally, our model is able to accurately capture sudden changes in density as can be seen with the voids in Figure \ref{visual_plot}.

The limitations of the HOD model are further illustrated by our results in Figure \ref{power_plot}, which presents various summary statistics of the fields. As shown, the score-based generative model produces samples with power spectra that accurately match the true hydro-simulation across all relevant scales. In contrast, the HOD model is under confident across all scales with a bias at lower k modes. The cross-correlation displayed in Figure \ref{power_plot} also highlights our model's ability to interpolate positional differences between high density regions in N-body and hydro-simulations. The cross-correlation is sensitive to the position of features in images as it captures their Fourier phase. It is known that halo positions slightly differ in N-body simulations compared to hydro-simulations due to the gravitational effects of the additional baryons. The HOD cannot correct for this difference in halo positions, however, the fact that our model's cross-correlation is higher than the HOD for all scales highlights its ability to correct this discrepancy.

Although the ideal cross-correlation is one across all scales, it is important to note that the decrease in cross-correlation at higher k modes for both the diffusion model samples and the HOD samples is expected, as non-linear processes dominate making it inherently more difficult in predicting the exact spatial distribution of galaxies solely from dark matter density fields. Regardless, in future work, we are interested in examining whether the score-generative model is able to extract the full information content available in dark matter N-body density fields for predicting the spatial distribution of galaxies.

While Figure \ref{power_plot} visually assesses the accuracy of the HOD model compared to the diffusion model for a single hydrodynamical simulation, Figure \ref{tarp_plot} quantitatively evaluates the accuracy of the power spectra across all 100 hydrodynamical simulation examples in the test set. Specifically, Figure \ref{tarp_plot} illustrates whether the power spectra of samples generated from the HOD and diffusion models accurately represent the distribution of possible power spectra of hydrodynamical galaxy count fields. Our diffusion model, which is closer to the diagonal line indicating perfect accuracy, demonstrates a significant improvement over the HOD model. However, our model still exhibits a bias at lower credibility values, indicating the need for further training improvements. This bias may result from the insufficient number of training examples (900 hydro-simulations) used in training the diffusion model. To address this, future work will explore data augmentation and pretraining on additional cosmological simulation datasets to enhance the model's generalization capabilities.

The primary objective of this work is to evaluate the potential of score-based generative models as reliable emulators of hydrodynamical fields, such as galaxy counts, at a fraction of the computational cost. Achieving this would facilitate cosmological inference tasks where many hydrodynamical simulations need to be generated to compare different models with observed galaxy distributions. By directly conditioning on the dark matter density field, this method also represents a significant step towards making forward models differentiable in cosmological simulations, completely removing the need to run a non-differentiable halo finder algorithm \citep{davis1985evolution, springel2001populating}. Consequently, to assess the viability of our diffusion model for cosmological inference, we aim to determine whether galaxy count samples generated by our diffusion model are sufficiently accurate to serve as training data for simulation-based inference methods to predict unbiased cosmological parameter values. Specifically, we will verify if the predictions derived from our model's samples are consistent with those obtained from real hydrodynamical simulations. This validation would demonstrate the potential of score-based generative models as an efficient method for producing training data of hydrodynamical simulations that are otherwise computationally prohibitive.

\section*{Acknowledgements}
This project was developed as part of the Simons Collaboration on ``Learning the Universe.'' The Flatiron Institute is supported by the Simons Foundation.  The work is in part supported by computational resources provided by Calcul Quebec and the Digital Research Alliance of Canada. A.B. and M.H. are supported by the Simons Collaboration on ``Learning the Universe.`` Y. H. and L. P.-L. acknowledge support from the Canada Research Chairs Program, the Natural Sciences and Engineering Research Council of Canada (NSERC) through grants RGPIN- 2020-05073 and 05102, and the Fonds de recherche du Qu\'ebec through grants 2022-NC-301305 and 300397. R.L. and A.A. are supported by NSERC CGS D scholarship.

\bibliography{citations}
\bibliographystyle{icml2024}

\newpage
\appendix
\onecolumn
\section{Appendix}
\label{sec:appendixA}
We present further summary statistics at different cosmologies to demonstrate our model's ability to generalize.
\vspace{-0.3cm}
\begin{figure}[h!]
    \centering
    \resizebox{0.9\textwidth}{!}{ 
        \begin{minipage}{\textwidth}
            \centering
            \begin{subfigure}[b]{0.32\linewidth}
                \includegraphics[width=\linewidth]{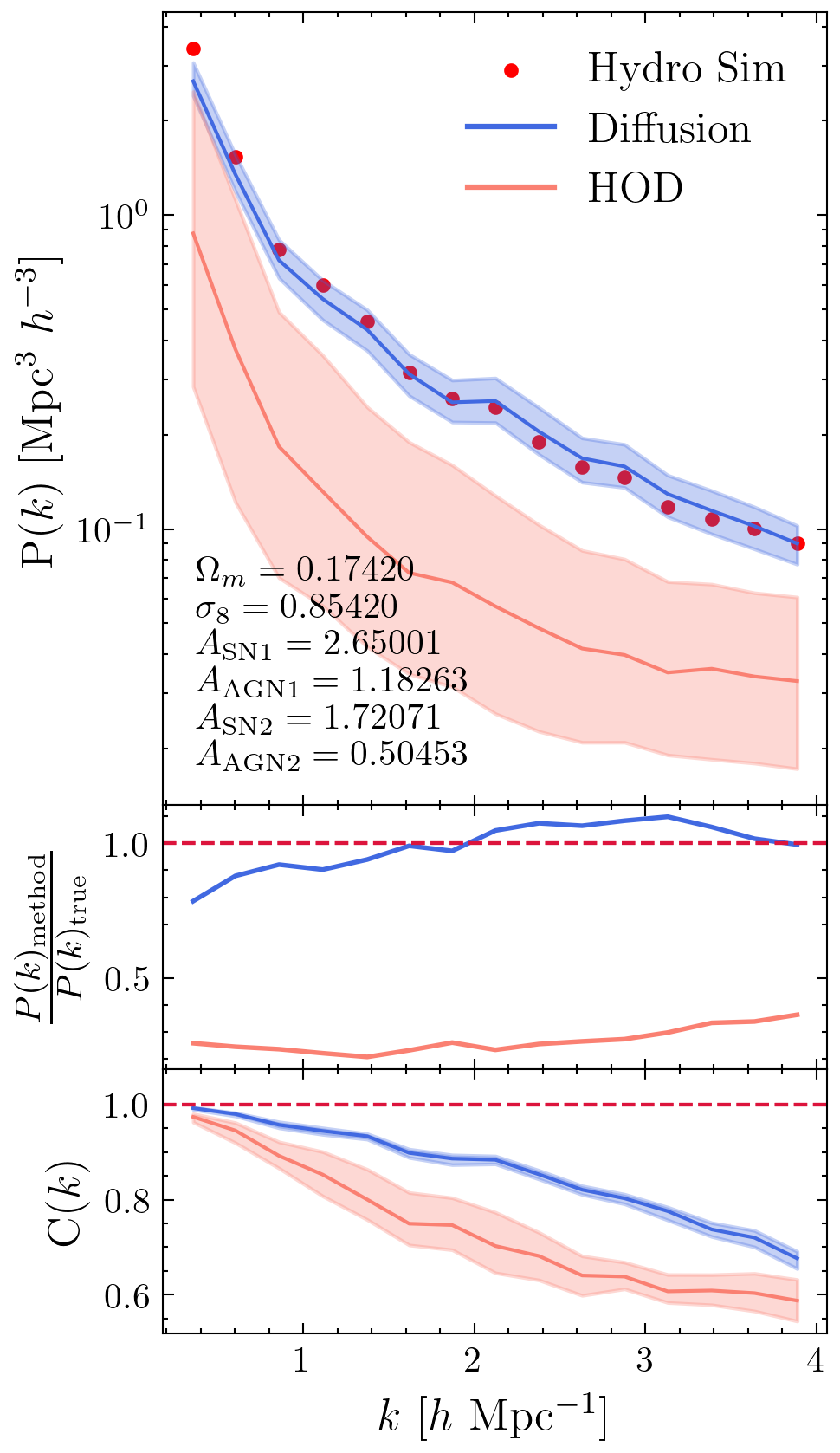}
                \caption{}
                \label{power_935}
            \end{subfigure}
            \hfill
            \begin{subfigure}[b]{0.32\linewidth}
                \includegraphics[width=\linewidth]{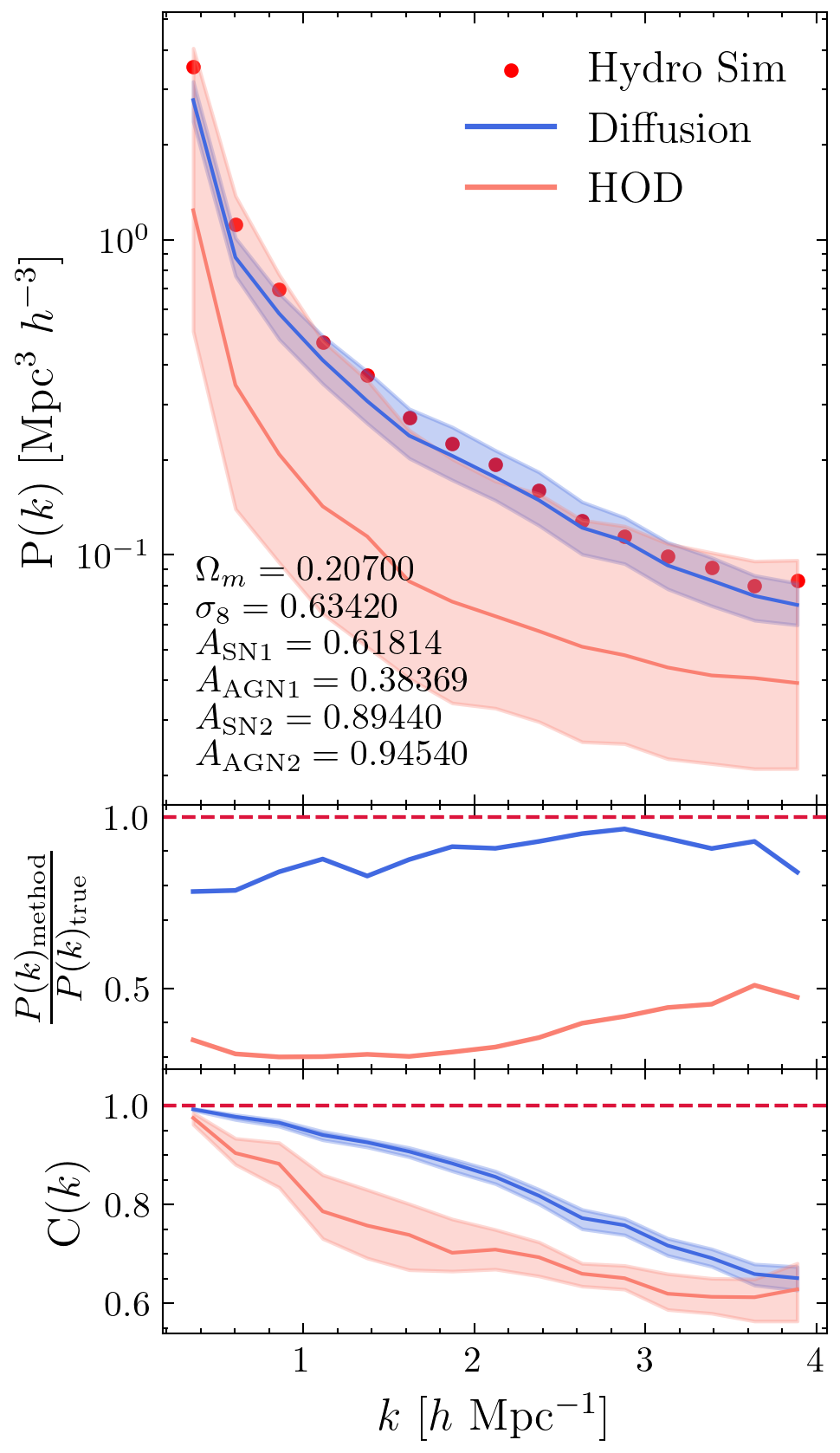}
                \caption{}
                \label{power_934}
            \end{subfigure}
            \hfill
            \begin{subfigure}[b]{0.32\linewidth}
                \includegraphics[width=\linewidth]{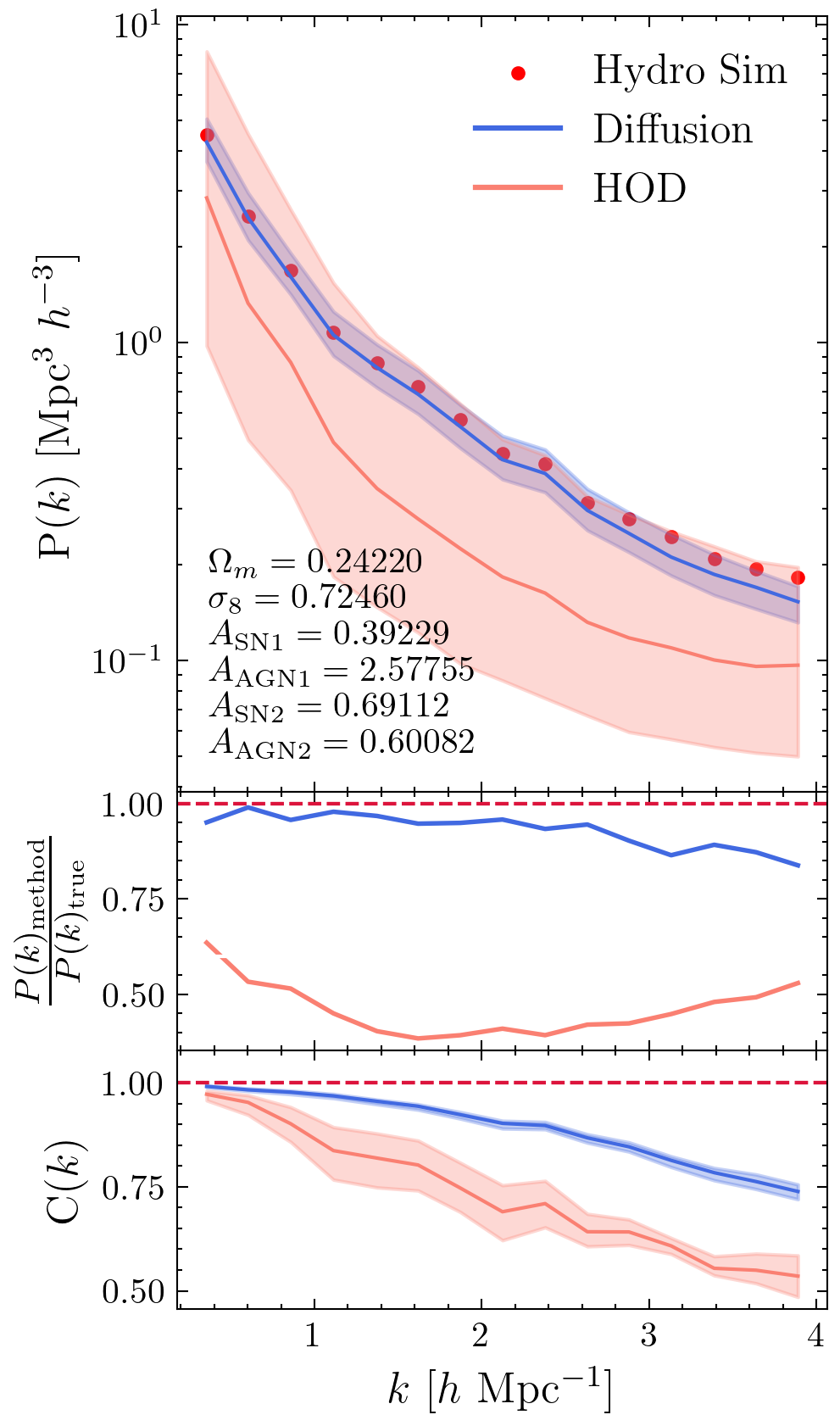}
                \caption{}
                \label{power_933}
            \end{subfigure}
            
            \medskip
            \begin{subfigure}[b]{0.32\linewidth}
                \includegraphics[width=\linewidth]{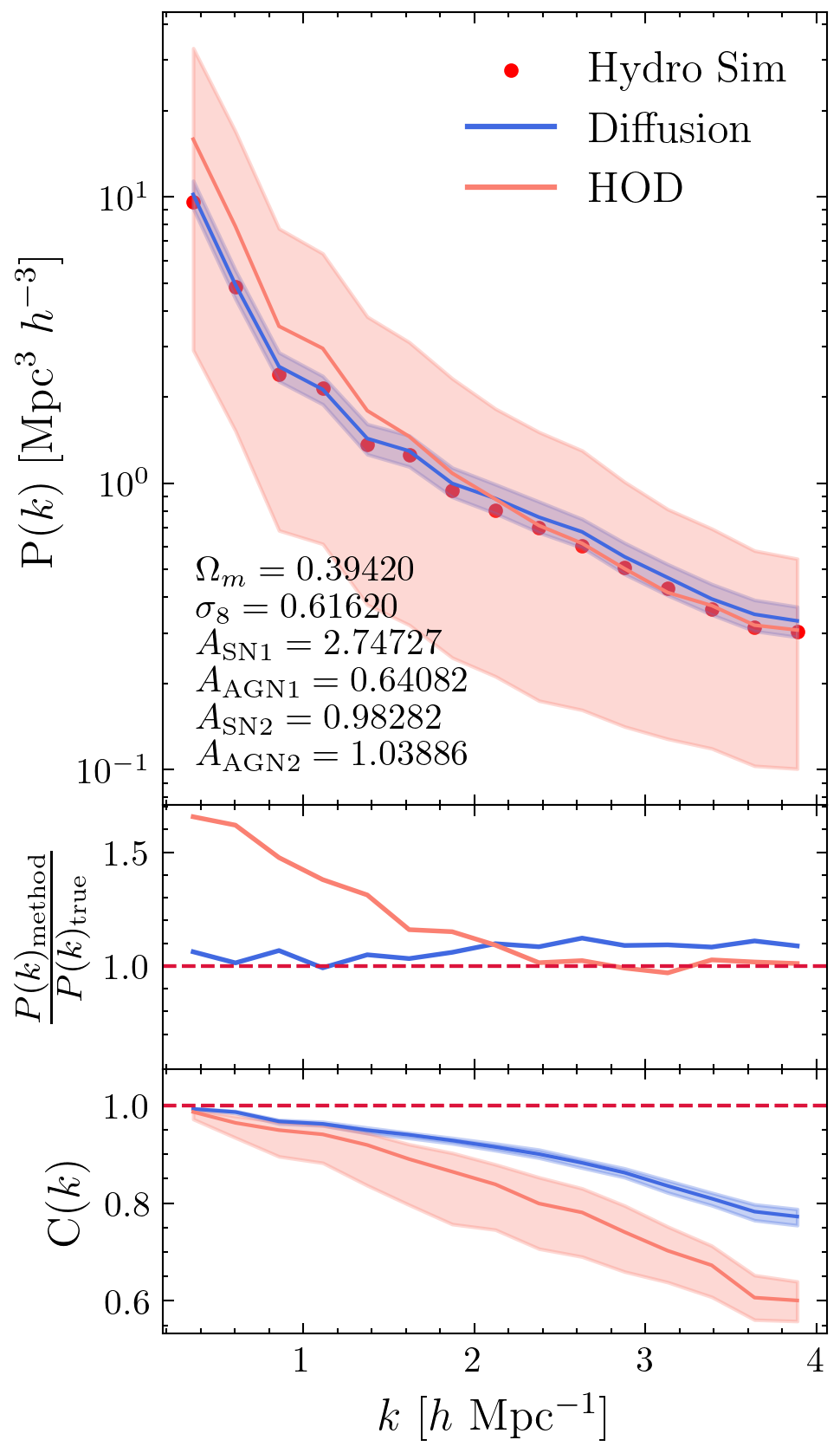}
                \caption{}
                \label{power_955}
            \end{subfigure}
            \hfill
            \begin{subfigure}[b]{0.32\linewidth}
                \includegraphics[width=\linewidth]{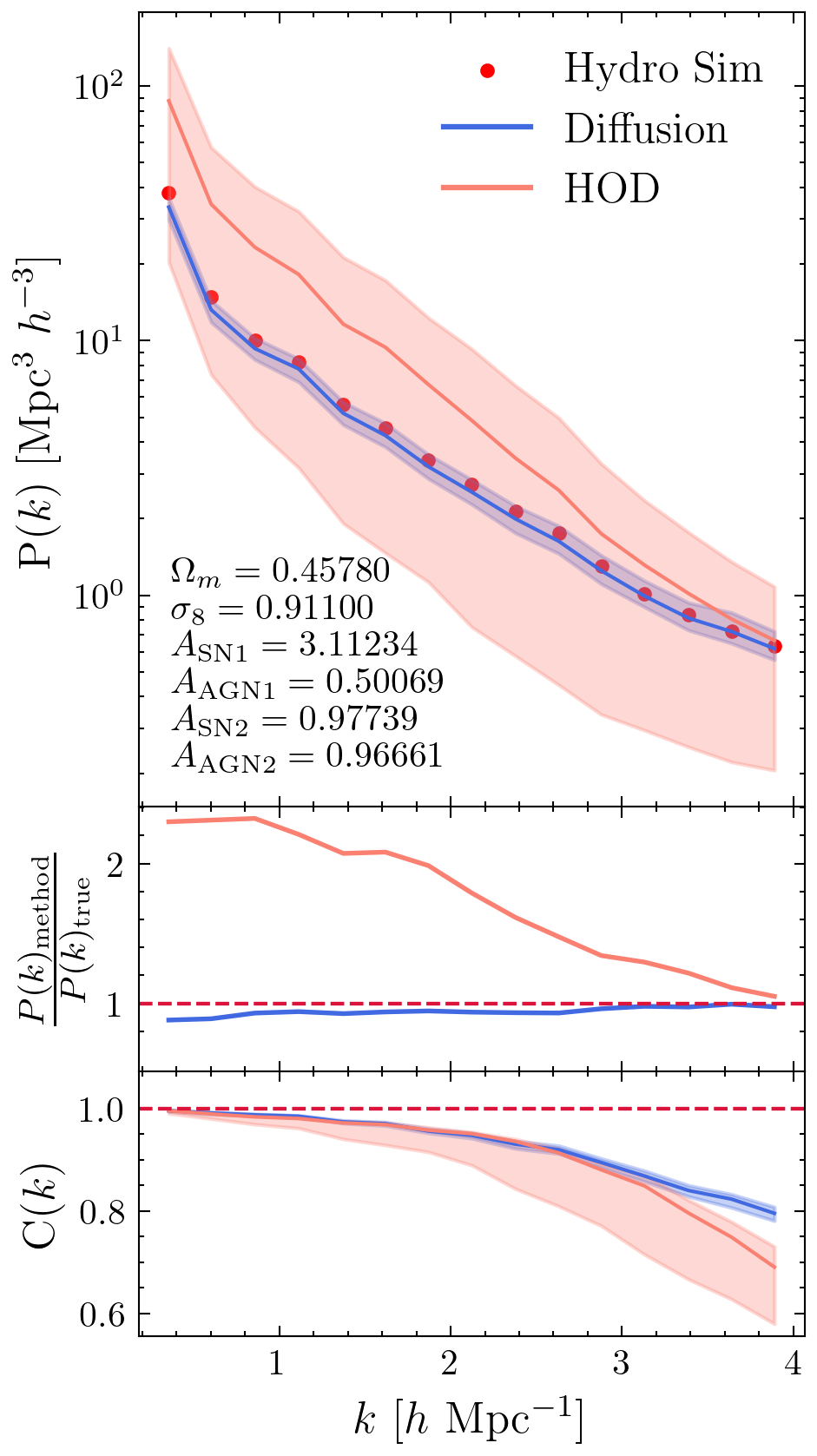}
                \caption{}
                \label{power_967}
            \end{subfigure}
            \hfill
            \begin{subfigure}[b]{0.32\linewidth}
                \includegraphics[width=\linewidth]{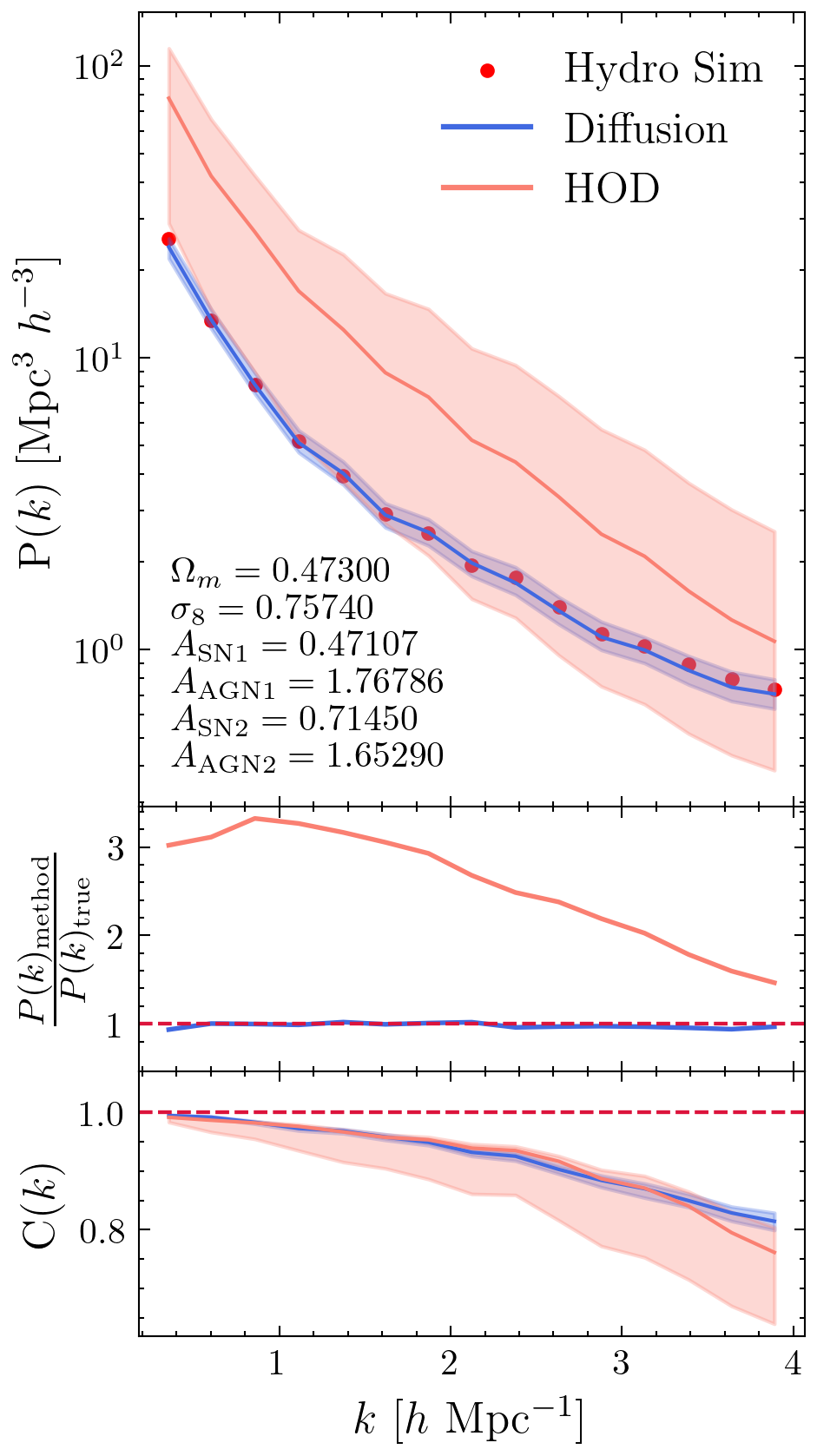}
                \caption{}
                \label{power_920}
            \end{subfigure}
        \end{minipage}
    }
    \caption{Summary statistics from various hydrodynamical simulations from the test set, encompassing a broad range of cosmological parameter values, are presented alongside results from samples generated using both the HOD method and the diffusion model. These results demonstrate that the diffusion model effectively generalizes across a wide range of cosmological parameter values beyond the training set. In contrast, the HOD method shows inferior generalization capabilities for simulations with different cosmological parameter values showcased here.}
    \label{fig:full_figure}
\end{figure}

\end{document}